\documentclass[aps,prb,twocolumn,superscriptaddress, show pacs]{revtex4-2}
\usepackage{bm, amsmath}
\usepackage{graphicx}
\usepackage{color}
\usepackage{epstopdf}
\usepackage{graphicx}
\usepackage{dcolumn}
\usepackage{bm}
\usepackage{subfigure}
\usepackage[rawfloats=true]{floatrow} 
\begin{document}

\title{Reorientation transition between square and hexagonal skyrmion lattices  near the saturation into the homogeneous state in quasi-two-dimensional chiral magnets}

\author{Andrey O. Leonov}
\thanks{Corresponding author: leonov@hiroshima-u.ac.jp}
\affiliation{Department of Chemistry, Faculty of Science, Hiroshima University Kagamiyama, Higashi Hiroshima, Hiroshima 739-8526, Japan}
\affiliation{International Institute for Sustainability with Knotted Chiral Meta Matter, Kagamiyama, Higashi Hiroshima, Hiroshima 739-8526, Japan}

\date{\today}

\begin{abstract}
{I revisit the well-known phase transition between the hexagonal skyrmion lattice and the homogeneous state within the phenomenological  Dzyaloshinskii theory for chiral magnets  which includes only the exchange, Dzyaloshinskii-Moriya and Zeeman energy contributions. 
I show that, in a narrow field range near the saturation field, the hexagonal skyrmion order gradually transforms into a square arrangement of skyrmions. Then, by the second-order phase transition during which the lattice period diverges, the square skyrmion lattice releases a set of repulsive isolated skyrmions. On decreasing magnetic field, isolated skyrmions re-condense into the square lattice at the same critical field as soon as their eigen-energy becomes negative with respect to the field-aligned state. 
The underlying reason of the  reorientation transition between two skyrmion orders can be deduced from the energy density distribution within isolated skyrmions surrounded by the homogeneous state. When the negative energy within the ring-shaped area at the skyrmion outskirt outweighs the positive energy amount around the skyrmion axis, skyrmions tend to form the square lattice, in which the overlap of skyrmion profiles results in smaller energy losses as compared with the hexagonal crystal. With the further decreasing field, the hexagonal lattice achieves smaller energy density in comparison with the square one due to the denser packing of individual skyrmions.
%
}
\end{abstract}

         
\maketitle

\section{Introduction}

\begin{figure*}
\includegraphics[width=1.5\columnwidth]{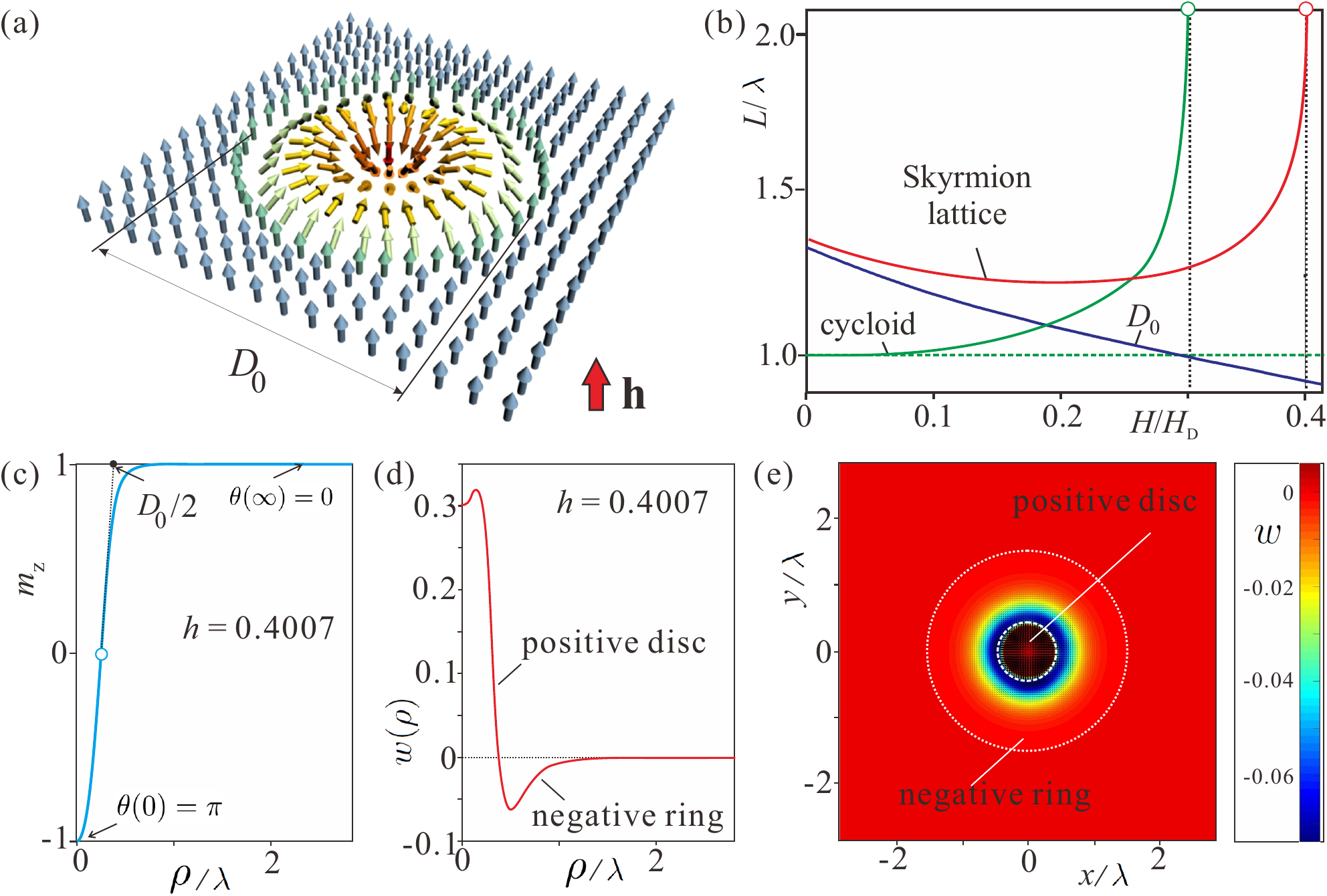}
\caption{(color online) 
\label{fig01} (a) Schematics of an isolated N\'eel skyrmion in polar magnets with the C$_{nv}$ symmetry (or in multilayers with the induced DMI (\ref{DMI})). (b) Equilibrium sizes of the skyrmion core $D_0$ (blue line) and the hexagonal SkL period (red line) compared with the corresponding period of the cycloid (green line) and showing expansion in an applied magnetic field.  (c) The localized magnetization profile $m_z(\rho)$ obtained by the cross-cut of a skyrmion through its center for $h=0.4007$. 
(d) The energy density $w(\rho)$ featuring parts with the negative and positive energy densities as calculated with respect to the homogeneous state at $h=0.4007$. (e) The corresponding color plot of the energy density on the plane $xy$. The length scale is measured in units of $\lambda$ (\ref{lambda}).}
\end{figure*}

Chiral magnetic skyrmions \cite{JETP89,Bogdanov94} are prominent solutions within the phenomenological theory (\ref{functional}) introduced by Dzyaloshinskii \cite{Dz64},  which are (i) localized, (ii) axisymmetric, and have (iii) fixed rotation sense of the magnetization (Fig. \ref{fig01} (a)).  
Surrounded by the homogeneously magnetized state, the relevant length scale of this magnetic inhomogeneity \cite{Bogdanov94,Le} is tuned by the competition between direct and chiral exchange (Dzyaloshinskii-Moriya interaction, DMI,  \cite{Dz58,Le11,Moriya,Dz64}).  In addition, DMI is an essential "ingredient" to overcome the constraints of the Hobart-Derrick theorem \cite{Hobart,Derrick} and to protect skyrmions from radial instability.

Historically, skyrmions were first experimentally identified to form A-phases near the ordering temperatures of bulk cubic helimagnets such as the itinerant magnets, MnSi \cite{Kadowaki,Muehlbauer09} and FeGe \cite{FeGe}, and the Mott insulator, Cu${_2}$OSeO${_3}$ \cite{Seki2012,Weiler}. 
Soon after, chiral skyrmions have been  microscopically observed in thin layers of cubic helimagnets (Fe,Co)Si \cite{yuFeCoSi} and FeGe \cite{yuFeGe} in a broad range of temperatures and magnetic fields far below the precursor regions.
Presently, versatile multilayer structures are considered as perfect systems to host skyrmions: the breaking of the inversion symmetry and the induced DMI originate from the interfaces between a heavy metal and the skyrmion-hosting magnetic layers as occurs, e.g., in  PdFe/Ir (111) \cite{Romming2013}.
These artificial material systems represent a 2D arena for the N\'eel skyrmions  (Fig. \ref{fig01} (a)), in which the magnetization rotates along the radial direction with zero helicity.

In modern spintronics, nanometer-size 2D skyrmions are considered as promising objects for the next generation memory and logic devices, which may be controlled and manipulated as information bits \cite{Fert2013,Tomasello14,APL2016}.
On the other hand, skyrmionic "particles" may also be driven together to form complex non-collinear magnetic textures – skyrmion lattices (SkL) \cite{yuFeCoSi,yuFeGe}. 

According to the commonly accepted paradigm \cite{Bogdanov94,Mukai22},  the mechanism of lattice formation through nucleation and condensation of isolated skyrmions follows a classification introduced by DeGennes for (continuous) transitions into incommensurate modulated phases and is dubbed nucleation-type phase transition \cite{deGennes}.
At some critical field, the eigen-energy of an isolated skyrmion becomes negative with respect to the surrounding homogeneous state. As a result, skyrmions tend to tile the whole plane with some equilibrium inter-skyrmion spacing.
Obviously, the phase transition relies on the discontinuous creation of skyrmions (read for details Refs. \onlinecite{Bogdanov94,IS} as well as the classification by de Gennes \cite{deGennes}).
The formation of an SkL is determined by the stability of the localized solitonic cores and their geometrical incompatibility that frustrates homogeneous space-filling. 
In the previos works \cite{Bogdanov94,Le11,Mukai22}, it was found  that a hexagonal skyrmion ordering has the lowest energy due to the densest packing of individual skyrmions in the whole field range within the basic Dzyaloshinskii model (\ref{functional}).

On the other hand, with the increasing magnetic field,  
the hexagonal skyrmion lattice transforms into the homogeneous state by infinite expansion of the period at the same critical field (red line in Fig. \ref{fig01} (b)). By this process, the lattice releases a set of repulsive isolated skyrmions \cite{Bogdanov94,Le11}. This transition excludes the formation of coexisting states -- SkL and the homogeneous state. 
The skyrmion-skyrmion interaction bears a repulsive character because of a fixed sense of the magnetization rotation \cite{repulsion}.
The distorted one-dimensional cycloids also infinitely expand their period \cite{Togawa} and transform into a system of isolated $2\pi$ domain walls separating domains with
the magnetization along the applied field (green line in Fig. \ref{fig01} (b)).

\begin{figure*}
\includegraphics[width=1.5\columnwidth]{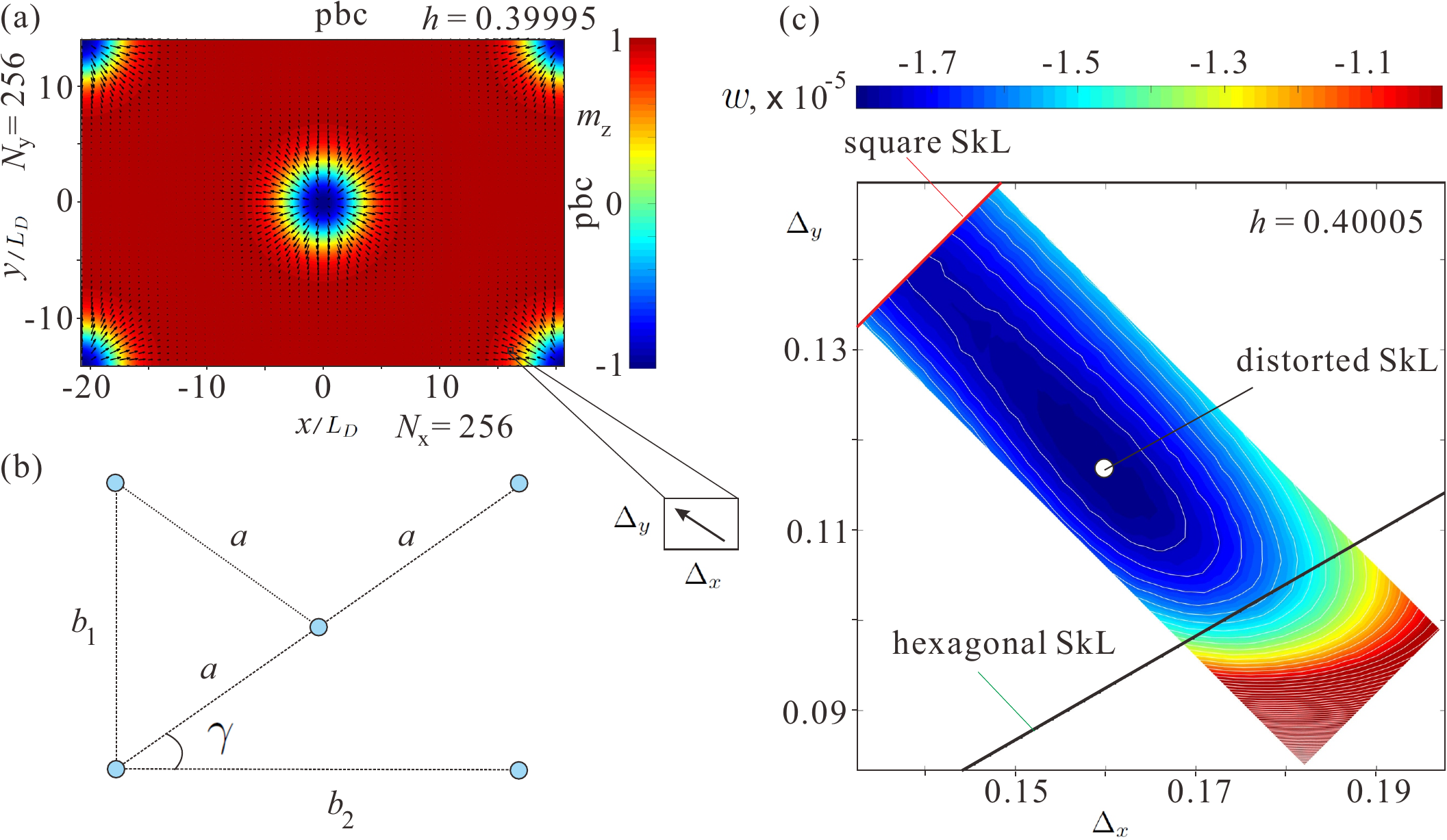}
\caption{(color online) (a) Schematics of a computational unit cell corresponding to a distorted SkL. The distribution of the magnetization within skyrmions retains its axisymmetric circular shape. The number of discretization points is equal along $x$ and $y$, $N_x=256, \, N_y=256$. The cell sizes, on the contrary, are varied to search for a deformed SkL with the lowest energy density.  (b) The characteristic geometric parameters of the unit cell, which exhibit the inter-skyrmion distances $b_1, b_2$, and $a$ as well as the chracteristic angle $\gamma$. (c)  The energy density of distorted SkLs computed by integration of Eq. (\ref{functional}) for different values of lattice spacings and for $h=0.40005$. The well-discernible energy minimum is formed for the skyrmion ordering, which is transient from the square (red line) to the hexagonal (black line) skyrmion arrangement.  The length scale is measured in units of $L_D$ (\ref{units}).
\label{fig02}}
\end{figure*}

In the present manuscript, I re-examine the described phase transition between an equilibrium hexagonal SkL and the field-polarized homogeneous state. 
I show that, although the above-mentioned arguments on the nucleation-type phase transition are well-founded, it is however the square skyrmion lattice, in which isolated skyrmions condense first. The square skyrmion arrangement represents the global energy minimum of the system. It allows to achieve the space tiling by skyrmions and at the same time to minimize the energy losses within the overlapped skyrmion profiles.  With the decreasing magnetic field, prompted to reach the densest space filling, the square SkL deforms and gives rise to two types of distorted SkLs. In a narrow field range, the distorted SkLs represent the global minima of the system, which eventually turn into two hexagonal SkLs. This reorientation transition was overlooked in previous studies presumably due to a small field interval. Moreover, if the simulations are performed for a skyrmion unit cell preserving its hexagonal symmetry, one would see an energy minimum for the hexagonal skyrmion order up to its saturation into the homogeneous state. The present numerical results reveal that the hexagonal SkL is not an energy minimum (not even a local one) for these field values.
Hence, the findings of the present manuscript strongly change the picture of the formation and evolution of skyrmion orderings on the verge with the homogeneous state.

\section{Phenomenological theory} 

The magnetic energy density of a quasi-two-dimensional chiral ferromagnet 
in the simplest isotropic form can be written as the sum of the exchange, the DMI, and the Zeeman energy density contributions, correspondingly \cite{Dz64}:
\begin{equation}
w(\mathbf{m})=\sum_{i,j}(\partial_i m_j)^2+w_{DMI}-\mathbf{m}\cdot\mathbf{h}.
\label{functional}
\end{equation}
The functional (\ref{functional}) includes only basic interactions essential to stabilize versatile isolated and modulated inhomogeneous states such as spirals and SkLs as well as isolated skyrmions and kinks. 

The non-dimensional units have been introduced to make the results general and encompassing as well as to be directly mapped to any material system.
Spatial coordinates $\mathbf{x}$ are measured in units of the characteristic length of modulated states $L_D$. $A>0$ is the exchange stiffness, $D$ is the Dzyaloshinskii constant.
\begin{equation}
L_D=A/D, \mathbf{h} = \mathbf{H}/H_0, H_0 = D^2/A|\mathbf{M}|.
\label{units}
\end{equation}
$\mathbf{h}$ is the  magnetic field applied along $z$ axis.
The magnetization vector $\mathbf{m}(x,y)$ is normalized to unity.

The DMI energy density  has the following form specific for chiral magnets with the C$_{nv}$ symmetry (or with induced DMI, which has the same functional form):
\begin{equation}
w_{DMI} = m_x\partial_x m_z-m_z\partial_x m_x+m_y\partial_y m_z-m_z\partial_y m_y, 
\label{DMI}
\end{equation}
where $\partial_x=\partial/\partial x,\, \partial_y=\partial/\partial y$.

Alternatively, the length scale can be measured in units of $\lambda$:
\begin{equation}
\lambda=4\pi L_D,
\label{lambda}
\end{equation}
which is the period of the spiral state for zero magnetic field (e.g., 18 nm for the bulk MnSi or 60 nm for Cu$_2$OSeO$_3$ \cite{Seki2012,Crisanti}). 
In actual simulations, we measure the length in units of $L_D$ (\ref{units}). Dividing by $4 \pi$, we get the length scale in units of $\lambda$, which provides a direct comparison with a specific material system. 

\begin{figure*}
\includegraphics[width=1.6\columnwidth]{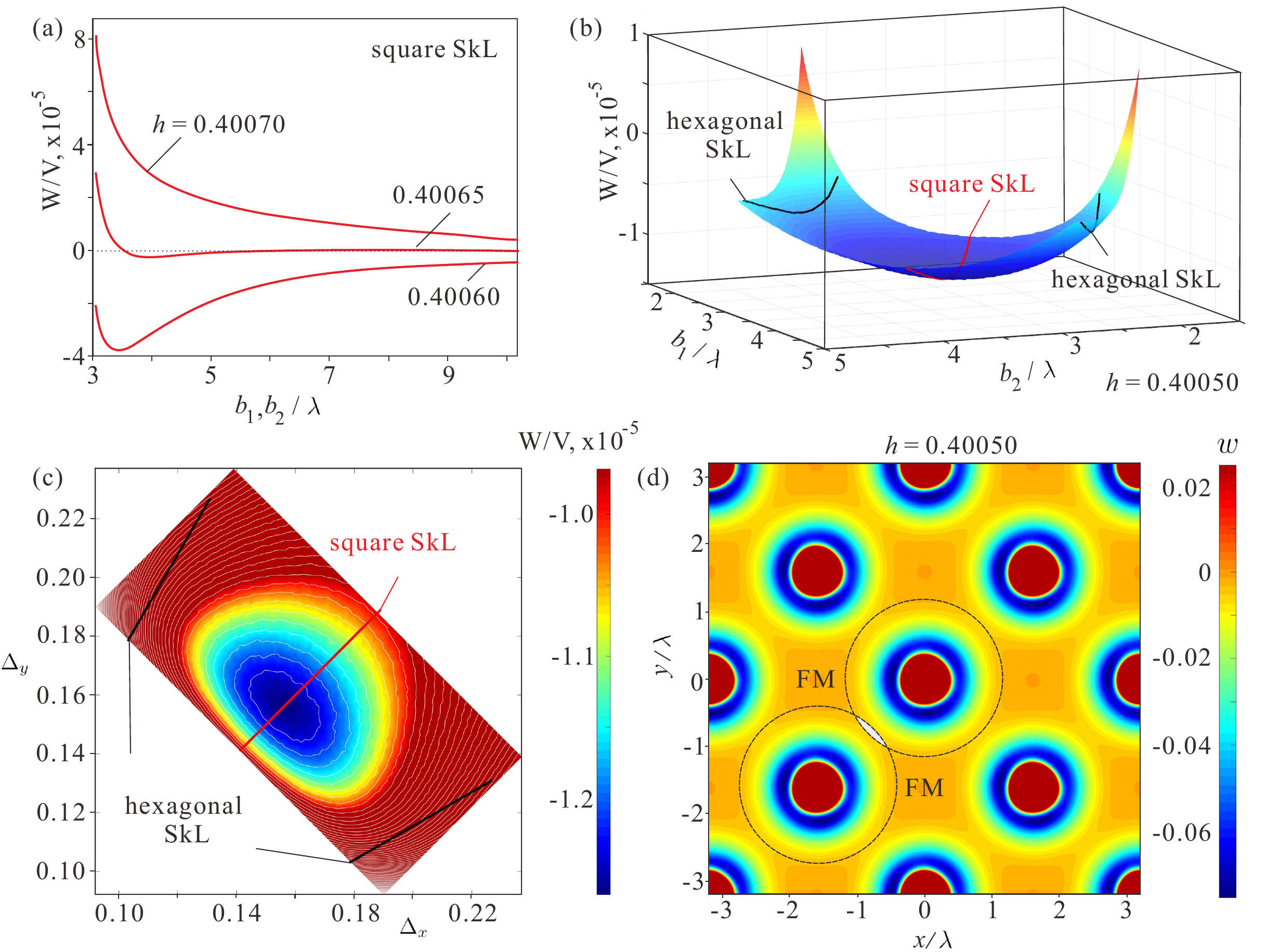}
\caption{(color online) (a) The energy density of the square SkL in dependence on the inter-skyrmion distance $b_1$ measured in units of $\lambda$ with respect to the homogeneous state. Below the threshold field $h_{\mathrm{IS}}$ (\ref{hIS}), there is a well-discernible energy minimum corresponding to an equilibrium lattice period. (b) The energy density of distorted skyrmion orders plotted as a surface in dependence on the parameters $b_1$ and $b_2$. The square SkL (highlighted by the red curve) reaches a global energy minimum. The parameters for the hexagonal SkLs correspond to black curves on a hill of the energy surface, i.e., the hexagonal SkL is not an energy minimum of the system. $h=0.40050$. (c) The color plot of the energy density on the plane of $\Delta_x$ and $\Delta_y$. This plot is simply a top view of (b), but with a zoomed range of the energy density near the minimum. Black and red lines show the parameters for the hexagonal and square SkLs, correspondingly. (d) The energy density distribution within the square SkL. Dashed circles represent overlapping skyrmion profiles with some energy loss within the gray-shaded regions. Ferromagnetic state retains in interstitial regions. 
\label{fig03}}
\end{figure*}

We consider a 2D film of a ferromagnetic material on the $xy$-plane using periodic boundary conditions (pbc). 
As a primary numerical tool to minimize the functional (\ref{functional}), we use MuMax3 software package (version 3.10) which calculates magnetization dynamics by solving the Landau-Lifshitz equation with finite difference discretization technique \cite{mumax3}.
To double-check the validity of obtained solutions, we also use our own numerical routines, which are explicitly described in, e.g., Ref. \onlinecite{metamorphoses} and hence will be omitted here. 
All structures are minimized on the grid $256\times 256$. 
To check the stability of different skyrmion orderings, we compute the energy density (\ref{functional}) for different ratios of the grid spacings $\Delta_y$ and $\Delta_x$ (cell sizes in mumax3, Fig. \ref{fig02} (a)). 
Such an approach is well justified, since the inter-skyrmion distances near the saturation field are relatively large, and the axisymmetric distribution of the magnetization within skyrmion cores is preserved.  Thus, varying lattice spacings lead to the rearrangement of the constituent isolated skyrmions spanning all possible lattice orders.

Fig. \ref{fig02} (a) shows the centered rectangular unit cell used for computations of skyrmion orderings.  
To characterize the degree of SkL deformations, we introduce the following lengths and angles consistent with the square and the hexagonal SkLs (blue circles correspond to skyrmion centers, Fig. \ref{fig02} (b)): (i) within the square SkL, $b_1=b_2=a$, $\gamma=45^{\circ}$; (ii) within the hexagonal SkL, $b_1=a=b_2/\sqrt{3}$, $\gamma=30^{\circ}$.

As an example, Fig. \ref{fig01} (c) shows the energy color plot depending on the cell sizes for $h=0.40005$. The red and black lines highlight the grid spacings for the square ($\Delta_x=\Delta_y$) and the hexagonal ($\Delta_x=\Delta_y\sqrt{3}$) skyrmion lattices. The energy minimum corresponds to a distorted SkL with the lattice parameters on their way from the square skyrmion alignment to the hexagonal one. $\Delta_x^{min}=0.158,\, \Delta_y^{min}=0.118$ what corresponds to $b_1=256\Delta_x^{min}/4\pi=3.22,\,b_2=2.4$ and $\gamma=36.75^{\circ}$.

The energy density is computed with respect to the homogeneous state and acquires small values near the field of the phase transition (see the legend in Fig. \ref{fig02} (c)). 

\begin{figure}
\includegraphics[width=0.99\columnwidth]{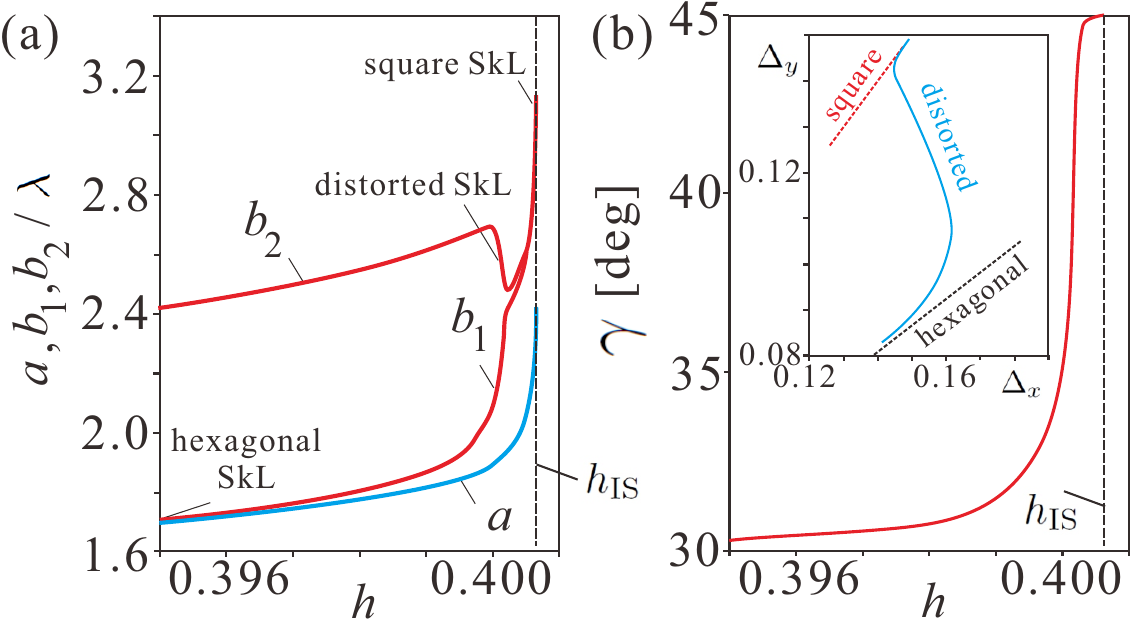}
\caption{(color online) (a)  Equilibrium parameters of the unit cell, as introduced in Fig. \ref{fig02} (b), in dependence on the field during the reorientation transition from the square into the hexagonal skyrmion arrangement. (b) The gradual field-driven change of the angle $\gamma$ from the value $45^{\circ}$ in the square SkL to the value $30^{\circ}$ within the hexagonal SkL. The inset shows corresponding cell sizes $\Delta_x$ and $\Delta_y$.
\label{fig04}}
\end{figure}

\section{Isolated skyrmions} 

Isolated skyrmions (Fig. \ref{fig01} (a)) can be thought of as static solitonic textures localized in two spatial directions. 
The magnetization in the center of skyrmion pointing opposite to an applied magnetic field rotates smoothly and reaches the state along the field at the outskirt of skyrmion. 
Conventionally, one uses spherical coordinates for the magnetization in isolated skyrmions,
\begin{equation}
\mathbf{m}= (\sin \theta(\rho) \cos \psi(\varphi),\sin \theta(\rho) \sin \psi(\varphi), \cos \theta(\rho)) 
\end{equation}
and cylindrical coordinates for the spatial variables \cite{Bogdanov94}, $\mathbf{r}=(\rho \cos \varphi, \rho \sin \varphi)$.
Therefore, the rotation of the magnetization in the isolated skyrmion is characterized by the dependence of the polar angle $\theta$ on the radial coordinate $\rho$ with the boundary conditions: $\theta(0)=\pi,\,\theta(\infty)=0$ (Fig. \ref{fig01} (c)). For the chosen C$_{nv}$ symmetry, $\psi(\varphi)=\varphi$.

The total energy of an isolated skyrmion with respect to the homogeneous state can be written as
\begin{align}
&W = \int\limits_0^{\infty} w (\rho) 2\pi\rho d \rho,\, \nonumber \\
&w(\rho) =  \left( \frac{d \theta}{d \rho } \right)^2 +\frac{\sin^2 \theta}{\rho^2} 
+ h\,(1- \cos \theta) + \frac{d \theta}{d \rho } + \frac{\sin 2 \theta}{2\rho}.
\label{energy1}
\end{align}
%

First, we examine the internal structure of such particle-like states when they possess the positive energy $W$ over the field-saturated state as realized for relatively strong magnetic fields. Skyrmion profiles $m_z(\rho)$ under these circumstances bear strongly localized character (Fig. \ref{fig01} (c)). Then, a skyrmion core diameter $D_0$ can be defined in analogy to definitions for domain wall width (the Lilley rule).  According to conventions in skyrmionics \cite{Bogdanov94,IS}, such arrow-like solutions can be decomposed into skyrmionic cores ("nuclei") and exponential "tails", which can be viewed as the "field" generated by the nucleus.

The energy density distribution $w(\rho)$ reveals two distinct regions: the positive energy disc is concentrated around the skyrmion center and is surrounded by the extended area with the negative energy density where the DMI dominates (Fig. \ref{fig01} (d)). The energy density is computed with respect to the homogeneous state and therefore reaches zero value at $\rho\rightarrow \infty$.
Two regions are highlighted by white dotted circles on a two-dimensional distribution of the energy density $w(x,y)$ (Fig. \ref{fig01} (e)). The color plot in Fig. \ref{fig01} (e) discerns the energy range from the minimal energy value to $w_{min}+0.08$, whereas in  Fig. \ref{fig01} (d) the whole energy diapason is shown. This allows to focus on the subtleties of the energy distribution at the skyrmion periphery.

The characteristic energy motif (Figs. \ref{fig01} (d), (e)) is know to cause the repulsive character of the inter-skyrmion potential \cite{Bogdanov94,repulsion}, since the approaching skyrmions would inevitably lose some amount of the negative energy density within the overlapping regions.

At the critical field \cite{Bogdanov94}
\begin{equation}
h_{\mathrm{IS}} = 0.400659,
\label{hIS}
\end{equation}
the negative energy density "accumulated" within the rings outweighs the positive contribution of the disc, i.e., the energy of an isolated skyrmion becomes
negative with respect to the surrounding homogeneous state below this field value and thus may initiate condensation into a skyrmion lattice. 

\begin{figure}
\includegraphics[width=0.8\columnwidth]{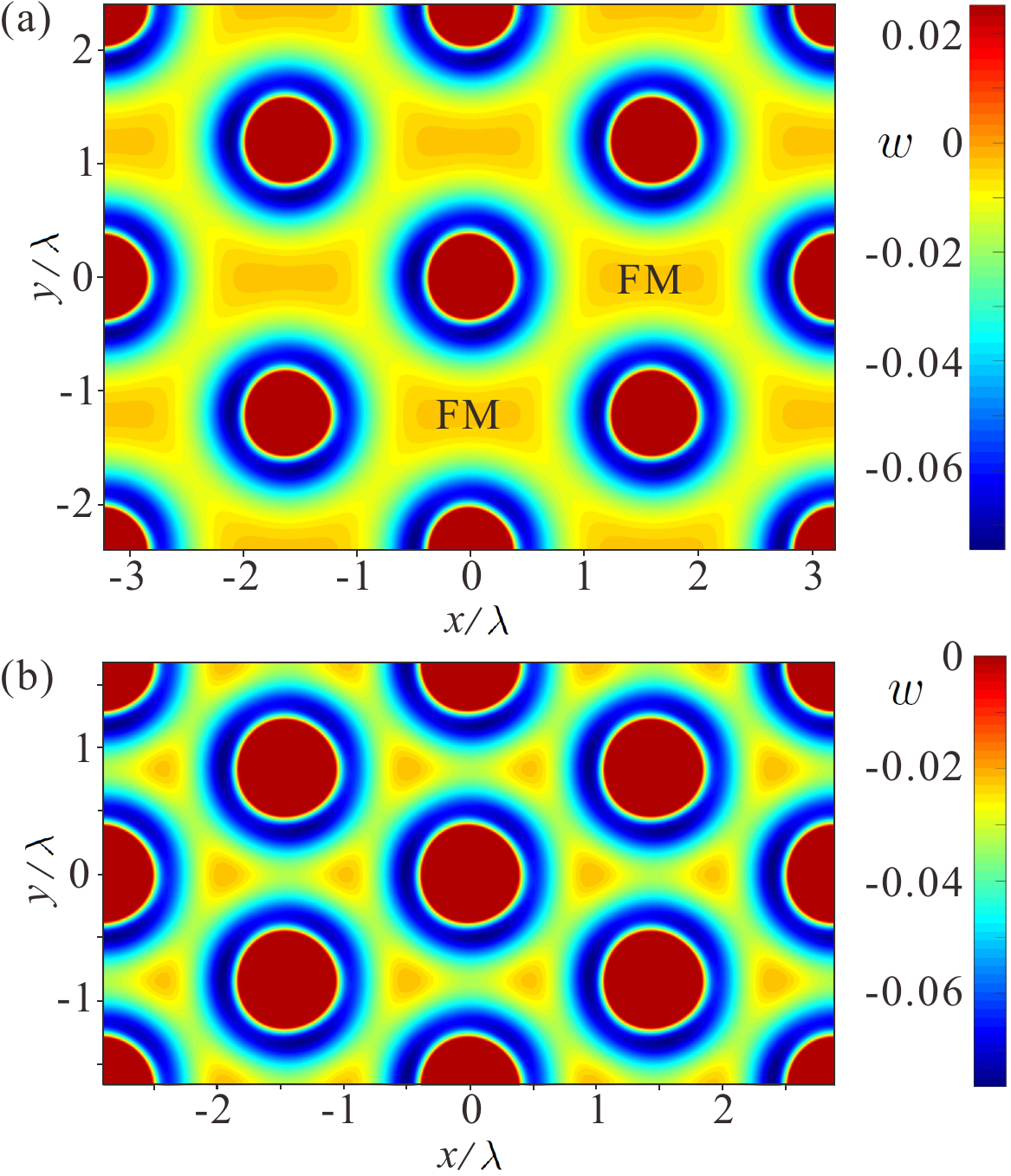}
\caption{(color online) Contour plots for the energy density distributions within the distorted (a) and the hexagonal SkLs (b) for $h=0.40005$ and $h=0.395$, correspondigly. In (a), the legend highlights the energy range $[w_{min},w_{min}+0.1]$ and thus indicates that the rectangular-shaped regions with the ferromagnetic state acquire almost zero energy density. In (b), the legend "zooms" the interval $[w_{min},0]$: the non-zero energy density within triangular regions becomes clearly discernible and thus indicates additional magnetization rotation around the central points with the upward magnetization.  
\label{fig05}}
\end{figure}

\section{Skyrmion orders} 

\subsection{Square SkLs}

Below the critical field $h_{\mathrm{IS}}$ (\ref{hIS}), the energy of a square skyrmion lattice has a minimum for some equilibrium inter-skyrmion distance.
Fig. \ref{fig03} (a) exhibits the energy densities of the square SkLs, 
\begin{equation}
\varepsilon = (1/V)\int w(x,y)dV, \,V=N_xN_y\Delta_x\Delta_y, \nonumber
\end{equation}
measured with respect to the homogeneous state for different values of the field. The integration is performed over the unit cell (Fig. \ref{fig02} (a)) with $V=N_xN_y\Delta_x\Delta_y$ being its volume. 

\begin{figure*}
\includegraphics[width=1.7\columnwidth]{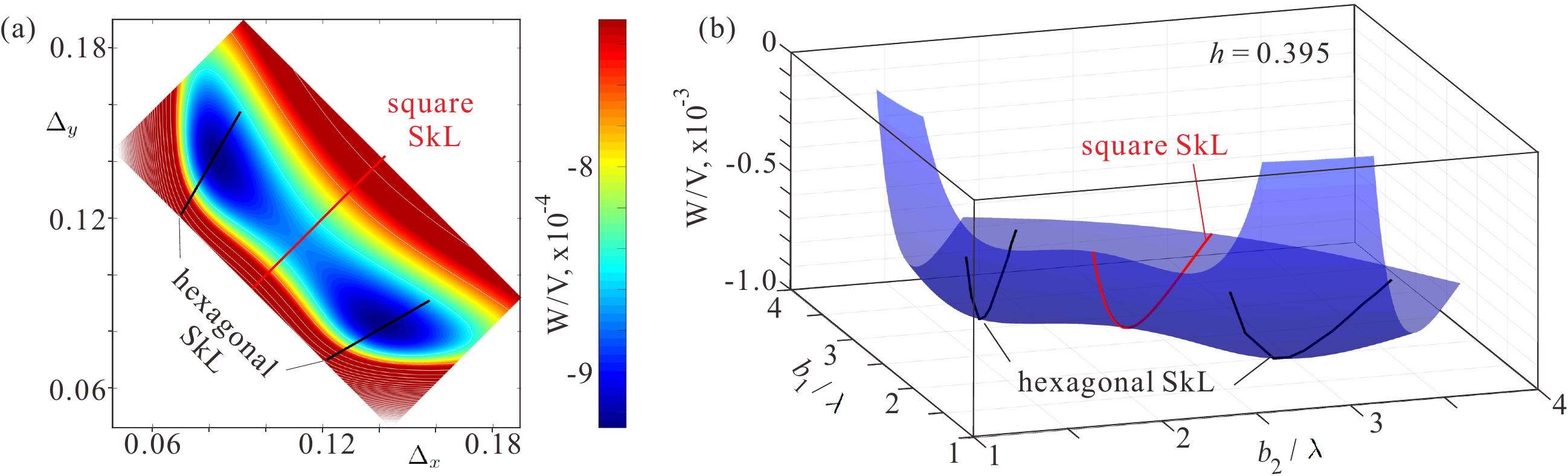}
\caption{(color online) (a) The color plot of the energy density on the plane of $\Delta_x$ and $\Delta_y$. This plot represents a top view of (b), but with a zoomed range of the energy density near the global energy minimum. Black and red lines show the parameters for the hexagonal and square SkLs, correspondingly. (b) The energy density of distorted skyrmion orders plotted as a surface in dependence on the parameters $b_1$ and $b_2$. The hexagonal SkLs (highlighted by the black curves) reach global energy minima. The parameters for the square SkL constitute a red curve with the minimum being a saddle point, i.e., there is no solution for the square SkL for this value of the field. 
\label{fig06}}
\end{figure*}

These energy curves represent 2D cross-cuts of the 3D energy surfaces plotted for variable lattice spacings $\Delta_x$ and $\Delta_y$ (Fig. \ref{fig03} (b)). The red and black curves indicate the energies of the square and the hexagonal skyrmion orders. As clearly seen, only the square SkL acquires the global energy minimum whereas the hexagonal SkLs are not minima at all. Fig. \ref{fig03} (c) shows the same energy plot as a top view with white contour lines indicating the energy levels. 
The color plot discerns the energy range from the minimal energy value to $\varepsilon_{min}+3\cdot 10^{-6}$ to make visible the topology of the energy surface in the direct vicinity of the energy minimum.

These findings can be readily explained by the aforementioned energy density distribution in Fig. \ref{fig01} (e). 
The square skyrmion ordering achieves smaller energy loss in the overlapping regions (highlighted by the dashed black lines and the gray-shaded regions) as compared with the hexagonal lattice (Fig. \ref{fig03} (d)). Obviously, this advantage is stipulated by four neighbors in the square ordering versus six neighbors -- in the hexagonal one. The square-shaped regions (orange-shaded) with the field-polarized state occupy the inter-skyrmion space. 


\subsection{Distorted skyrmion lattices} 

Fig. \ref{fig04} (a) shows the lattice parameters near the saturation field.

Below the critical field, the square SkL undergoes a reorientation transition into the hexagonal skyrmion order via intermediate distorted SkLs. 
An example of the oblique SkL for $h=0.39995$ is shown in Fig. \ref{fig02} (a). The square lattice corresponds to a saddle point between two oblique SkLs with slightly higher energy density as shown by the color plot in Fig. \ref{fig02} (c) for $h=0.40005$.

With the decreasing magnetic field, the unit cell size $b_2$ passes through the maximal value at $h\approx 0.3999$ and then gradually decreases. The parameter $b_1$ diminishes directly from its infinite value at $h_{\mathrm{IS}}$.
The angle $\gamma$ (Fig. \ref{fig04} (b)) exhibits the expected evolution from $45^{\circ}$ at $h_{\mathrm{IS}}$ to $30^{\circ}$ (the precise value of the field is defined by the accuracy of the simulations and is slightly less than 0.395).  We notice that the angle retains its maximal value in a small field range near $h_{\mathrm{IS}}$ and then drastically decreases.
Inset in Fig. \ref{fig04} (b) shows the cell sizes $\Delta_x$ and $\Delta_y$ during the reorientation process. The far right ploint of the blue curve corresponds to $h=0.40065$, the far left -- to $h=0.395$. These values are explicitly expressed for the reproducibility of the obtained results. Multiplied by $N_x/4\pi$ and $N_y/4\pi$, these values turn into $b_1$ and $b_2$, respectively.
 Square and hexagonal SkLs are shown as dotted lines since they do not exist in the chosen field range and are reached as the limiting cases of the distorted SkLs.

During the reorientation transition, the vast square-shaped regions with the field-polarized state, which occupy the inter-skyrmion space within the square SkL  (Fig. \ref{fig03} (d)), elongate  within distorted SkLs (Fig. \ref{fig05} (a)) and eventually  squeeze into triangular regions surrounding skyrmion cells in hexagonal SkLs (Fig. \ref{fig05} (b)). 
The corresponding legends indicate that the energy density is zero in these regions within the distorted SkL whereas it is negative in the hexagonal one. 
This is related to the additional rotation of the magnetization, which develops around these characteristic points with the upward magnetization. Accordingly, the maximal value of the DMI energy density gradually decreases from zero to finite negative values: $w_{DMI}^{max}\approx -6.6\times 10^{-4}$ ($h=0.40050$); $-3.8\times 10^{-3}$ ($h=0.40005$); $-2.2\times 10^{-2}$ ($h=0.39500$). 
Thus, one can speculate that the stability of the square SkL near the saturation field can be addressed using the particle picture of isolated skyrmions, which slightly overlap their magnetization profiles.  The density of the skyrmions within the hexagonal SkL, however, could be better understood from the point of view of waves \cite{Rosch}: separate skyrmions preserve axisymmetric distribution of the magnetization near the cell center while the overlap of solutions in the
inter-skyrmion regions becomes pronounced even in the direct vicinity of $h_{\mathrm{IS}}$. In Ref. \onlinecite{Muehlbauer09} a triple spin-spiral crystal was used as a skeleton for such a skyrmion lattice in cubic chiral magnets. However, the transformation process of a square SkL into an assembly of isolated skyrmions cannot be described by
the picture of a triple spin-spiral crystal.

Fig. \ref{fig06} (a) shows the energy color plot depending on the cell sizes for $h=0.395$. The more general energy surface is shown in Fig. \ref{fig06} (b).
The red and black lines, as before, highlight the lattice parameters for the square  and the hexagonal  skyrmion lattices. 
It is seen that the reorientation process has almost finished with two hexagonal SkLs being the global minima of the functional (\ref{functional}). The square SkL constitutes a saddle point between two hexagonal SkLs, although within two-dimensional simulations preserving the same length of $\Delta_{x,y}$ it would be mistakenly treated as a local minimum. 
$\Delta_x^{min}=0.141,\, \Delta_y^{min}=0.083$ what corresponds to $b_1=2.8724,\,b_2=1.69$ and $\gamma=30.48^{\circ}$ and thus demonstrates slight lattice distortion. Two hexagonal lattices are rotated by the angle $\pi/6$ with respect to each other.

\section{Conclusions}

In the present paper, I addressed the process of skyrmion ordering into an extended skyrmion lattice as occurs at the critical field of saturation into the homogeneous state. I used the basic Dzyaloshinskii model, which includes only exchange, Dzyaloshinskii-Moriya and Zeeman energy terms. The model is isotropic and does not feature any anisotropy with its easy (or hard) axes in the plane $xy$, which would define skyrmion arrangement. 

Obviously, a well-defined skyrmion order could be favored by additional anisotropic contributions, e.g., by the exchange or cubic anisotropies. 
In particular, a high sensitivity of skyrmion order to the local magnetic anisotropy was reported in Ref. \onlinecite{Nakajima} for metastable skyrmions obtained at lower temperatures by thermal quenching in MnSi. 
The quenched skyrmions were shown to undergo a triangular-to-square lattice transition with decreasing magnetic field.  
Moreover, square and rhombic lattices of magnetic skyrmions were recently identified in centrosymmetric rare-earth compounds, such as Gd$_2$PdSi$_3$ and GdRu$_2$Si$_2$ \cite{Takagi}. 
Recently, formation of the triangular skyrmion lattice was found in a tetragonal polar magnet VOSe$_2$O$_5$ \cite{Kurumaji2017}. Adjacent to this phase, another magnetic phase of an incommensurate spin texture is identified at lower temperatures, tentatively assigned to a square skyrmion-lattice phase.

The results of the present paper on the skyrmion ordering, however, are based on entirely different principles and reveal distorted skyrmion lattices with the limiting cases of the hexagonal and square orders as global energy minima of the phenomenological functional (\ref{functional}).

At first, it is not obvious which skyrmion order prevails. On the one hand, the highest spatial density of skyrmions is achieved in the hexagonal SkL. On the other hand, the loss of the negative energy density in the overlapping regions is reduced in the square SkL.  By sorting out all possible skyrmion orders (which in general
represent rhombic Bravais lattices) near the transition into the homogeneous state, I found that isolated skyrmions first condense into the square SkL, which ensures the second objective to preserve the negative eigen-energy of isolated skyrmions as much as possible. With the decreasing magnetic field, the square SkL starts to deform and reaches the hexagonal skyrmion arrangement within a narrow field interval. This reorientation is driven by the tendency to satisfy the first condition of the closest skyrmion packing. I remark that at each field value only one skyrmion lattice corresponds to the energy minimum, i.e., the co-existence of the hexagonal and the square SkLs, which would represent local and global minima, is excluded. When the square SkL is realized, the hexagonal one is not even a local minimum. When the hexagonal order is finally reached, the square SkL is a saddle point.

\textit{Acknowledgements. }
The author is grateful to Natsuki Mukai, Kaito Nakamura and Takayuki Shigenaga for useful discussions. 

\end{document}